\documentstyle[12pt,epsfig]{article}
\def\ls{\mathrel{\raise1.16pt\hbox{$<$}\kern-7.0pt %
\lower3.06pt\hbox{{$\scriptstyle \sim$}}}}         %

\begin{document}

\hrule
\vskip 0.2in
\Large{The Complete Star Formation History of the Universe}

\normalsize \author{Alan Heavens$^1$, Benjamin Panter$^1$, Raul
Jimenez$^2$, James Dunlop$^1$}

Alan Heavens$^1$, Benjamin
Panter$^1$, Raul Jimenez$^2$, James Dunlop$^1$

\noindent
$^1$Institute for Astronomy, University of Edinburgh, Blackford Hill,
Edinburgh EH9-3HJ, UK; afh, bdp, jsd@roe.ac.uk\\
$^2$Department of Physics \& Astronomy, University of
Pennsylvania, 209 South 33rd Street, Philadelphia, PA 19104-6396,
USA; raulj@physics.upenn.edu \vskip 0.2in \hrule \vskip 0.2in

\noindent {\bf The determination of the star-formation history of
the Universe is a key goal of modern cosmology, as it is crucial
to our understanding of how structure in the Universe forms and
evolves. A picture has built up over recent years, piece-by-piece,
by observing young stars in distant galaxies at different times in
the
%past\cite{Gallego95,Lilly96,Steidel96,Connolly97,Tresse98,Glazebrook99,Cowie99,Sullivan00,Scott02,Chapman03,Ouchi03,Stanway03}.
past\cite{Gallego95}-\cite{Stanway03}.
 These studies indicated that the stellar birthrate peaked
some 8 billion years ago, and then declined by a factor of around
ten to its present value. Here we report on a new study which
obtains the complete star formation history by analysing the
fossil record of the stellar populations of 96545 nearby galaxies.
Broadly, our results support those derived from high-redshift
galaxies elsewhere in the Universe.  We find, however, that the
peak of star formation was more recent - around 5 billion years
ago. Our study also shows that the bigger the stellar mass of the
galaxy, the earlier the stars were formed. This striking result
indicates a very different formation history for high- and
low-mass formation.}

%Introduction

The optical stellar spectrum of a galaxy can be used as a probe of
both its past star formation history, and the metallicity of its
gas as a function of time.  The spectra of a large sample of
nearby galaxies therefore acts as a useful fossil record of the
star formation rate (SFR) of the Universe, allowing estimates of
the SFR over a very wide range of cosmic time, from the same
internally consistent data set, and with very small statistical
errors.  The alternative is to use the finite speed of light to
view galaxies at different cosmic epochs, and to look for signs of
recent star formation.  The present method decouples the star
formation from the mass assembly - it takes into account all stars
which end up in normal galaxies today, and it is much less
sensitive to the large corrections, for extinction and for
unobserved small galaxies, which must be applied to high-redshift
studies.

%\section{Survey details and methods}

The spectral data used in this analysis come from the Sloan
Digital Sky Survey data release 1 (SDSS DR1), from an area of 1360
square degrees.  The main sample has red apparent magnitude limits
of $15.0 \le m_R \le 17.77$, and we also place a cut on surface
brightness of $\mu_R<23.0$ (see \cite{Shen03} for discussion of
this).  The redshift range is $0.005 < z <0.34$, with a median of
0.1.  This leaves 96545 galaxies in this study. Full details of
the SDSS are available at {\tt http://www.sdss.org/}. The spectra
are top-hat smoothed to 20\AA\, resolution, for comparison with
the models of Jimenez et al\cite{Jimenez03}, and emission-line
regions are removed, as these are generally non-stellar and
outside the scope of the theoretical model employed.

%\subsection{MOPED}

The speed of MOPED means we are not restricted to assuming simple
parametrizations of the star formation
history\cite{Baldry02,Glazebrook03}.  For each spectrum, we
recover the mass of stars created in 11 time periods, which are
mostly equally spaced logarithmically in look-back time, separated
by factors of 2.07, but with the first boundary at a redshift of
2. We assume a Salpeter initial mass function, and a cosmology
given by the best-fitting parameters determined by the WMAP
satellite\cite{Spergel03}: $\Omega_m=0.27$, $\Omega_v=0.73$, $H_0
= 71\, $km$\,$s$^{-1}$Mpc$^{-1}$.  For each time period, we also
recover the average metallicity of the gas, and an overall dust
parameter for the galaxy, assuming a Large Magellanic Cloud
extinction curve\cite{Gordon03}.  Thus we have a 23-dimensional
parameter space to search.  A straightforward maximum likelihood
solution using the full spectrum of all the SDSS galaxies is
impractical, so we use the patented radical lossless data
compression algorithm MOPED \cite{HJL00} to compress each spectrum
to 23 numbers, and we find the set of 23 parameters which fit
these MOPED coefficients most accurately.  The massive data
compression allows much faster determination of the parameters and
the errors on them, and, importantly, can be shown to be lossless
in ideal cases, in the sense that the error bars are not increased
by using the MOPED coefficients rather than the entire spectrum.
More details are given in papers developing and testing
MOPED\cite{HJL00,Reichardt01,Panter03}.

With this method, we obtain the mass of stars created in each galaxy
$g$ in each period of time (relative to the time of emission of the
galaxy light), $\delta M_{*g}(t)$.  After redistribution of the stars
to a set of time bins fixed in cosmic time, we estimate the star
formation rate per unit co-moving volume by $\dot\rho_*(t) = \sum_g
\delta M_{*g}/(V_{\rm max} \delta t)$, where $\delta t$ is the width
of the time bin.  $V_{\rm max}$ is the maximum volume in which the
galaxy could be placed and still contribute to the star formation rate
estimate at time $t$.  Since we have no knowledge of the star
formation rate at redshifts less than the observed redshift of the
galaxy, this introduces a redshift cutoff.  Other limits come from the
magnitude and surface brightness limits of the sample.  These are
calculated by computing the galaxy spectrum as a function of time
(from the recovered SFR parameters) and computing the expected
observed $R-$band flux and surface brightness.  We take account of the
3 arc sec fibre aperture by correcting the spectrum upwards by the
ratio of the flux in $R$ in a Petrosian radius (determined by the
photometry) to the fibre flux.  For individual galaxies this is likely
to fail, but for the population as a whole there is evidence from the
colours \cite{Glazebrook03} that fibre-placing is such that there is
no significant systematic difference in the sampling of stellar populations by
the spectroscopy and photometry.

%{Results}

\begin{figure*}
\begin{center}
\includegraphics[width=10cm,height=14cm,angle=270]{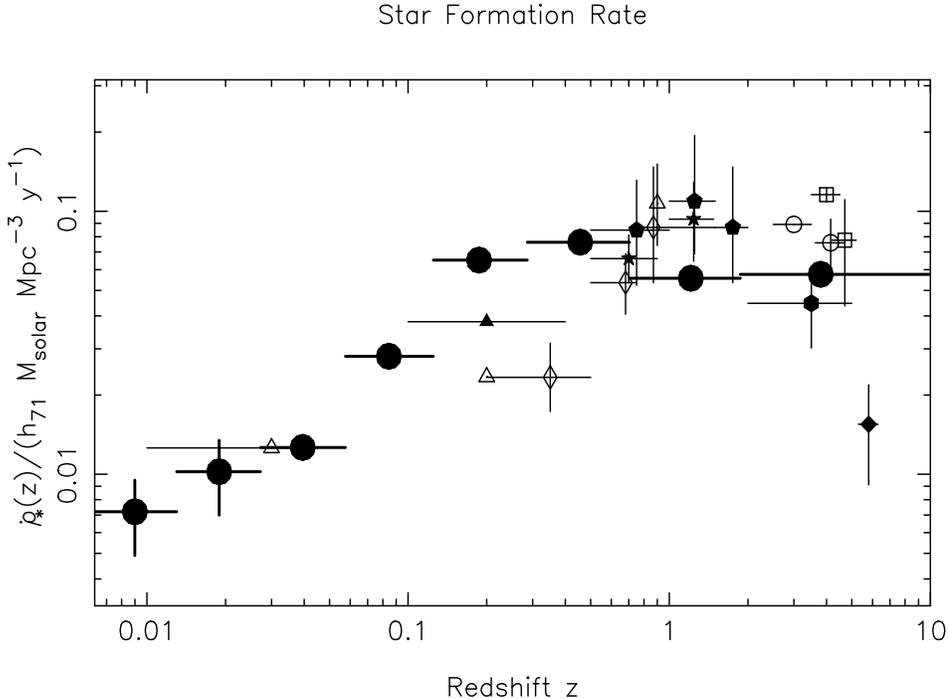}
\caption{{\small The star formation history of the Universe. The
star formation rate recovered from the ``fossil record" in the
SDSS is shown by the eight large filled circles. The horizontal
error bars represent the size of the bin in redshift. Vertical
errors are bootstrap errors, and are invisibly small for most
bins. The other symbols correspond to independent determinations
using instantaneous measurements of the star formation rate, as
follows; $H\alpha$ measurements are open triangles at $z\simeq
0.03$\cite{Gallego95}, $z\simeq 0.2$\cite{Tresse98}, $z\simeq
0.9$\cite{Glazebrook99}; UV from Subaru\cite{Ouchi03} (open
squares), GOODS\cite{Stanway03} (filled diamond), HST
etc\cite{Steidel96} (open circles), CFRS\cite{Lilly96} (open
diamonds), HDF\cite{Connolly97} (filled pentagons),
galaxies\cite{Cowie99} (stars), galaxies\cite{Sullivan00} (filled
triangle). The filled hexagon at $z=3.5$ represents a new estimate
of the star-formation density provided by sub-mm galaxies in the
redshift range $2 < z < 5$. This was derived by integrating the
sub-mm source number counts\cite{Scott02} down to $S_{850\mu m} =
1$ mJy, and assuming that 75\% of such sources lie at $z > 2$ (in
line with recent redshift measurements\cite{Chapman03}). In
general the agreement is very good. However, there are two
important results in our study which result from the extremely
small vertical error bars. First, we find that 26\% of the mass of
stars in the present-day Universe were formed at $z > 2$. Second,
while we confirm the previous measurements of a generally high
level of star-formation activity at $z \simeq 1$, we find that
global star-formation density peaked at significantly lower
redshifts than previously claimed, in the bin spanning the
redshift range $0.3 < z < 0.8$. The reason for this difference is
clarified in Figure 2. $h_{71}$ is the Hubble constant in units of
71 km$\,$s$^{-1}\, $Mpc$^{-1}$.}}
\end{center}
\end{figure*}

In Fig. 1 we show the co-moving star formation rate determined by
MOPED and SDSS DR1, as a function of redshift, along with results
from other determinations, largely based on instantaneous star
formation rate indicators (UV flux, H$\alpha$ emission, sub-mm
emission etc). At a basic level the new data show good agreement
over the redshift range of $0.01 < z < 6$.  The low-redshift
decline of SFR is clearly seen, and the SFR appears to have
flattened. Moreover, the level of star-formation activity inferred
at high-redshift ($z > 2$) is in rather good agreement with the
completely independent values inferred from observations of
high-redshift galaxies, suggesting that the dust corrections made
in such studies are reasonable, if a little overestimated. The
broad agreement of the SFR determined locally from the fossil
record of the SDSS with the determinations from high-redshift
galaxies gives support for the Copernican Principle, i.e. that the
Earth has no special location in the Universe.  It also supports
the current standard cosmological model, as the volume elements
assumed in the high-redshift studies are sensitive to the
cosmological parameters. Note, however, that this does assume that
our assumption about the initial mass function is reasonable, as
we are inferring the number of early-forming high-mass stars (now
dead) from the numbers of early-forming low-mass stars, which are
still contributing to the galaxy spectrum.

One of our main results is that the period of star formation is
broader than previously thought, and occurs at a lower redshift
$z\simeq 0.6$, rather than 1 or more.  Specifically, we find that
26\% of the mass of stars in the present-day universe was formed
at $z > 2$ (cf \cite{Dickinson03}). The average metallicity rises
from 0.44 (relative to Solar) at high $z$ to a peak of 0.8 at
$z\simeq 1$ before declining to a level around 0.25 at the present
time.

As we explain below, we believe our result differs because it
includes the contributions made by all galaxies over a very wide
mass range, extending down to galaxies with $L \sim 2\times
10^{-3}L_*$.  Note that virtually all 96545 galaxies contribute to
the $z>0.3$ bins, so the statistical errors (bootstrap estimates)
are negligible in comparison with modelling uncertainties and
residual uncertainties in flux calibration.  We also note that
because we are not dominated by statistical errors, the errors are
smaller in this approach than by analysing, for example, some
appropriately-weighted average of the spectra themselves.
Supplementary Information Figure 1 shows that parameter recovery
is robust.

Our second major result is apparent when we present the star
formation history of the sample divided into different ranges of
observed stellar mass.  We see strikingly in Fig. 2 that the
redshift at which star-formation activity peaks is an essentially
monotonically increasing function of final stellar mass.
Star-formation activity in the galaxies in the lowest mass bins
($M_*\simeq 10^{10} M_\odot$) peaked at $z \simeq 0.2$, while for
galaxies an order of magnitude more massive the peak lies at $z
\simeq 0.5$. At still higher masses, galaxies with masses
comparable to a present-day $L_*$ galaxy appear to have
experienced a peak in activity at $z \simeq 0.8$, while the
highest-mass systems ($M_*> 10^{12} M_\odot$) show a monotonic
decline in SFR in our data, with any peak constrained to lie at $z
> 2$. This provides a natural explanation for why the most massive
star-forming systems, such as the luminous sub-mm selected
galaxies, should be largely found to lie at high-redshift ($z >
2$;\cite{Chapman03}) while at the same time providing further
evidence that the bright sub-mm galaxies are indeed the
progenitors of today's massive ellipticals\cite{Dunlop01}.  The
importance of low-mass systems in low-redshift star formation has
been noted for example by \cite{PerezGonzalez03} and
\cite{Fujita03}.

\begin{figure*}
\begin{center}
\includegraphics[width=10cm,height=14cm,angle=270]{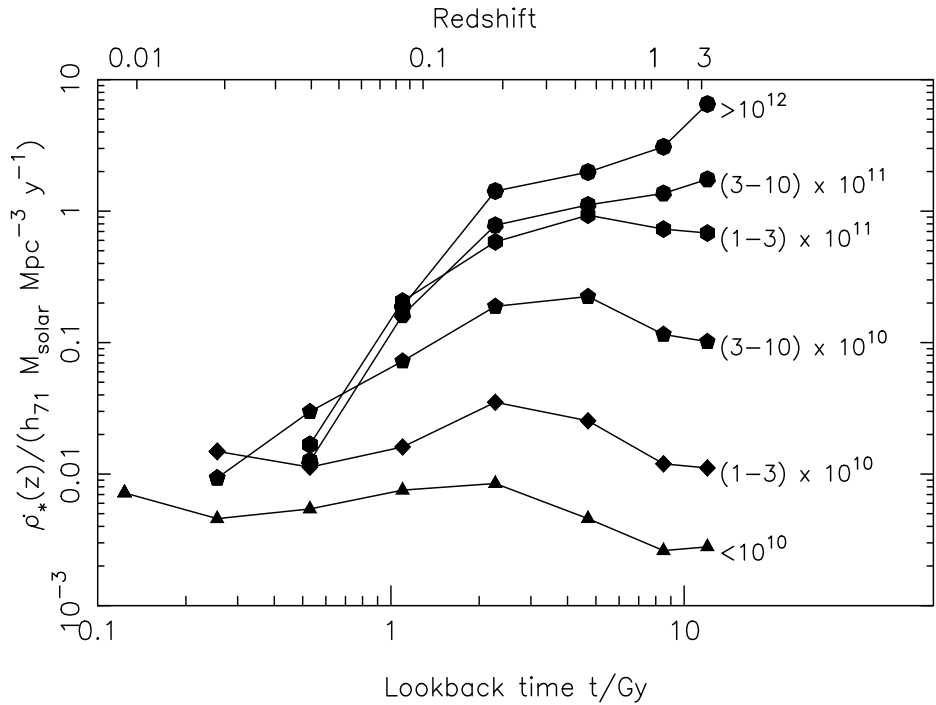}
\caption{{\small The star formation rate as a function of the
observed stellar mass of the galaxy.  Labels are in Solar Masses.
For clarity, the curves are offset vertically successively by 0.5
in the log, except for the most massive galaxies, which are offset
by an additional 1.0.  Note the clear trend for galaxies with
larger present-day stellar mass to have formed their stars
earlier.  The bulk of the star formation rate at $z \simeq 0.5$
comes from galaxies with present-day stellar masses in the range
$3-30 \times 10^{10}M_\odot$.  Note that the graph makes no
statement about when the masses were aggregated. To further test
the robustness of our findings we have reconstructed the star
formation history changing the dust model and the theoretical
stellar populations models. The shape of the star formation
history is hardly changed by changing the dust model to the
Calzetti extinction law\cite{Calzetti97}, or by using
Bruzual-Charlot\cite{BC93} stellar population models; the latter
allows us to assess the effect of systematic errors in our
determination of the SFR, since the Jimenez and Bruzual-Charlot
models are based on different stellar interior and atmospheres
models.}}
\end{center}
\end{figure*}

Indeed, the strong mass-dependence of the star formation history
provides a natural explanation of the high redshift of peak star
formation activity seen in other surveys, since they are sensitive
to the most massive objects only. The fact that we have now
discovered that global star-formation activity in fact peaks at
rather modest redshift is due to the fact that the peak epoch of
star-formation activity in objects of lower (present-day stellar)
masses ($< 3 \times 10^{11}M_\odot$) was at $z \ls 0.5$, and that
such lower-mass galaxies make a significant contribution to the
overall star-formation density. Given present-day observational
capabilities, this result could only be revealed by a method such
as used here, due to the fact that the fossil record approach
allows as to explore the star-formation history of galaxies
spanning over two decades in mass.  Supplementary information
Figure 2 demonstrates consistent results from volume-limited
subsamples.

Finally, we note that the mass dependence of peak star-formation
epoch revealed in Fig. 2 appears to mirror the mass dependence of
black-hole activity as recently seen in redshift surveys of both
radio-selected\cite{Waddington02} and X-ray
selected\cite{Hasinger03} active galactic nuclei.  Such apparently
anti-hierarchical behaviour (``downsizing'') is in fact quite
consistent with the standard cosmological model, in which galaxies
form in small units and merge - our method makes no statement
about whether the stellar mass at high redshift was in smaller
units or not. The behaviour we see is based on the {\em
present-day} stellar mass of the galaxies, and generally we would
expect more massive systems to be part of large-scale
overdensities, whose first star formation would occur earlier.
Furthermore, these results suggest a very different formation
history of low- and high-mass systems, as the mass assembly
proceeds in the opposite direction to the star formation.  This
will be explored in a separate paper.

Correspondence and requests for materials should be sent to Alan
Heavens (afh@roe.ac.uk)

{\bf Acknowledgments} We are grateful to Max Pettini, Masami Ouchi
and the referees for helpful remarks.  The SDSS is managed by the
Astrophysical Research Consortium (ARC) for the Participating
Institutions.
%
% Cut here if too long
%
The Participating Institutions are The University of Chicago,
Fermilab, the Institute for Advanced Study, the Japan
Participation Group, The Johns Hopkins University, Los Alamos
National Laboratory, the Max-Planck-Institute for Astronomy
(MPIA), the Max-Planck-Institute for Astrophysics (MPA), New
Mexico State University, University of Pittsburgh, Princeton
University, the United States Naval Observatory and the University
of Washington.

Supplementary information accompanies the paper on
www.nature.com/nature.

\end{document}